\documentclass[prb,floatfix,twocolumn,showpacs]{revtex4}
\usepackage{graphicx}
\usepackage{amsmath2000}
\usepackage{amsfonts}
\usepackage{bm}
\usepackage{latexsym}
\begin{document}
\title{The effects of magnetic field on the $d$-density wave order in the cuprates}
\author{Hoang K. Nguyen and Sudip Chakravarty}
\email[First author: ]{hknguyen@physics.ucla.edu}
\email[second author: ]{sudip@physics.ucla.edu}

\affiliation{Department of Physics and Astronomy, University of California Los
Angeles, Los Angeles, California 90095}
\date{\today}
\begin{abstract}
We consider the effects of a perpendicular magnetic field on the $d$-density wave order and conclude
that if the pseudogap phase in the cuprates is due to this order, then it is highly
insensitive to the magnetic field in the underdoped regime, while its sensitivity increases as the gap
vanishes in the overdoped regime. This appears to be consistent with the available experiments and can
be tested further in neutron scattering experiments. We also investigate the nature of the de Haas- van
Alphen effect in the ordered state and discuss the possibility of observing it.   
\end{abstract}
\maketitle

Recently it was argued\cite{Chakravarty1,Chakravarty2} that the observed pseudogap
in the cuprate superconductors
is due to the development of a second
type of order, the $d$-density wave order (DDW), which is a particle-hole condensate with an
internal ``angular momentum"  2.
The notion of  this broken symmetry  has allowed us to 
understand an array of experiments, although 
{\em a priori} there is
no sense in which such a state is close
to, or adiabatically continuable to, a
superconductor,  or  a Fermi liquid, which are the two most prominent states of matter.

From considerations of the simplest Hartree-Fock Hamiltonian that captures the broken symmetry of
the DDW and the electromagnetic gauge invariance, we study  the dependence of the DDW order parameter
as a  function of the applied  magnetic field perpendicular to CuO-planes. We find that the effect of
the magnetic field on the pseudogap is  rather weak, if the DDW order is well-developed. For example,
to destroy a DDW gap of magnitude of order 20 meV would require fields of order 1000 T, or
larger. Moreover, this gap is highly insensitive to fields of order 25 T. This is, of course, 
a characteristic of the underdoped regime. In the overdoped regime, where the DDW gap drops rapidly to
zero, the dependence on the magnetic field could be substantial.  

 We suggest that the recent neutron
scattering experiments, which report a tantalizing evidence of  DDW order\cite{Mook} in the form of a
resolution-limited elastic Bragg peak at the in-plane wave vector $(\pi/a,\pi/a)$, where $a$ is the
lattice spacing, should be carefully examined for its dependence on a perpendicular magnetic field. The
prediction is that the Bragg scattering intensity will be hardly affected, if the zero temperature  DDW
gap is substantial.\cite{Mook2} 

The effect of magnetic 
field on a superconductor is well known. Since the Cooper pairs
involve time reversed mates, the effect of any time reversal breaking perturbation
is very important. In contrast, the particle-hole pairs of the same spin orientation form the DDW
condensate, and, as a consequence, such a condensate is inherently  less affected by a magnetic field.

The contrasting response of  a
$d$-wave superconductor (DSC) and a DDW to applied magnetic field is further illuminated by examining
the nodal quasiparticles of these two systems. The nodal quasiparticles of a DSC do not form Landau
levels,\cite{Tesanovic} while those of a DDW state do.\cite{Nersesyan} This implies that, in principle,
it is possible to observe de Haas-van Alphen (dHvA) oscillations in a DDW state, but not in a $d$-wave
superconductor. The reason is that quasiparticles in a superconductor do not couple simply to the vector
potential
$\bf A$ corresponding to the magnetic field ${\mathbf B}={\bm \nabla}\times{\mathbf A}$, but to
the supercurrent
$(\bm{\nabla}
\varphi - 2e{\mathbf A}/\hbar c)$, where
$\varphi$ is the phase of the superconducting order parameter, $e$ is the electronic
charge, and $c$ is the velocity of light. We calculate dHvA oscillations of the magnetization in the
DDW state and find that although $\Delta B/B$ could be of reasonable magnitude, the amplitude of the
magnetization oscillations is weak, generically of  order 0.01$\mu_B$. Thus, observation of these
oscillations may be difficult.

Consider the mean field  Hamiltonian of a $d$-density wave  subjected to a magnetic field
applied perpendicular to the plane. With the choice of the Landau gauge, the Peierls substituted
Hamiltonian is written as:
\begin{eqnarray}
H=&-&\mu\sum_{m,n}c^+_{m,n}c_{mn}-t\sum_{m,n}\left(c^+_{m,n}c_{m+1,n}+\text{h.c}\right)\nonumber\\ 
&-&t\sum_{m,n}\left(e^{2\pi i\alpha m}c^+_{m,n}c_{m,n+1}+ \text{h. c.}\right) \nonumber\\
&-&\sum_{m,n}(-1)^{m+n}\Delta\left(-ic^+_{m,n}c_{m+1,n}+\text{h.c}\right) \nonumber \\
&+&\sum_{m,n}(-1)^{m+n}\Delta\left(-ie^{2\pi i\alpha m}c^+_{m,n}c_{m,n+1}+\text{h.
c.}\right)\nonumber\\ 
\label{eq:Hamiltonian}
\end{eqnarray}
Here $m$ and $n$ are the site labels of a square lattice in the $x$- and $y$-directions; 
$\alpha=\phi/\phi_0$ is the magnetic flux through each plaquette of the square lattice in terms of the
flux quantum
$\phi_0=hc/e$, and $c_{m,n}$ is the fermion destruction operator at the site $(m,n)$. Here on, the
hopping matrix element
$t$ will be set to unity. The DDW gap satisfies the self-consistency condition:
\begin{equation}
\Delta = -i(-1)^{m+n}V\langle c^+_{m,n}c_{m+1,n}-\text{h. c.} \rangle, 
\label{eq:gap_x}
\end{equation}
where the angular brackets denote the groundstate average. Due to the conservation of current, ensured by
the gauge invariance of  the Hamiltonian (\ref{eq:Hamiltonian}), the DDW gap could as well be:
\begin{equation}
\Delta =i(-1)^{m+n}V\langle e^{2\pi i\alpha m}c^+_{m,n}c_{m,n+1} -\text{h. c.}\rangle .
\label{eq:gap_y}
\end{equation}
Since a \textit{uniform} applied magnetic field cannot break translational invariance, all physical
quantities must be translationally invariant. Here
$V$ is an energy parameter that controls the strength of DDW pairing.

We rewrite the Hamiltonian in terms of the Fourier transformed variables, given by (with lattice
constant set to unity) 
\begin{equation}
c_{m,n} = \frac{1}{\sqrt{N}}\sum_{k_x,k_y}c_{k_x,k_y}e^{i(k_x m+ k_y n)},
\end{equation}
where $N$ is the total number of sites on the lattice, to get
\begin{eqnarray}
H &=& -\mu\sum_{k_x,k_y} c^\dagger_{k_x,k_y}c_{k_x,k_y} - 2\sum_{k_x,k_y}
\cos k_x c^\dagger_{k_x,k_y}c_{k_x,k_y} \nonumber \\
&-& \sum_{k_x,k_y} \left(e^{ik_y} c^\dagger_{k_x+2\pi\alpha,k_y}c_{k_x,k_y} + \text {h.
c.}\right)\nonumber \\
&+& \Delta\sum_{k_x,k_y} \left(ie^{ik_x} c^\dagger_{k_x+\pi,k_y+\pi}c_{k_x,k_y} + \text {h.
c.}\right)\nonumber \\
&-& \Delta\sum_{k_x,k_y} \left(ie^{ik_y} c^\dagger_{k_x+\pi+2\pi\alpha,k_y+\pi}c_{k_x,k_y} + \text {h.
c.}\right). 
\end{eqnarray}
One sees immediately that the states $(k_x,k_y)$, $(k_x+2\pi\alpha,k_y)$, $(k_x+4\pi\alpha,k_y)$,
{\ldots}, $(k_x+\pi,k_y+\pi)$, $(k_x+\pi+2\pi\alpha,k_y+\pi)$,
$(k_x+\pi+4\pi\alpha,k_y+\pi)$, {\ldots} are coupled. For rational flux, $\alpha =p/q$, where $p$ and
$q$ are relative primes, the set of coupled states is finite, containing $2q$ elements. At this
point, the connection to the Hofstadter problem\cite{Hofstadter} becomes particularly obvious. The
Hamiltonian can be written as the following quadratic form:
\begin{equation}
H=\sum_{\bf k}\Psi_{\bf k}^\dagger (H_{\bf k} - \mu ) \Psi_{\bf k},
\end{equation}
where $k_x\in [0,2\pi\alpha)$ and $k_y \in [0,\pi)$. The row vector $\Psi_{\bf k}^T$ is given by
\begin{equation}
\begin{split}
\Psi_{\bf k}^T=
(&c_{k_x,k_y}, c_{k_x+2\pi\alpha,k_y}, c_{k_x+4\pi\alpha,k_y}, \ldots,
c_{k_x+2\pi\alpha(q-1),k_y},\nonumber \\
&c_{k_x+\pi,k_y+\pi}, c_{k_x+\pi+2\pi\alpha,k_y+\pi},c_{k_x+\pi+4\pi\alpha,k_y+\pi},\ldots,\nonumber
\\ &c_{k_x+\pi+2\pi\alpha(q-1),k_y+\pi}) ,
\end{split}
\end{equation}
\begin{widetext}
and the matrix $H_{\bf k}$ is 
\begin{equation}
\begin{vmatrix}-2\cos k_x & -e^{-ik_y} & 0 & \ldots & -e^{ik_y} &-2i\Delta\cos k_x &
i\Delta e^{-ik_y} & 0 & \ldots & i\Delta e^{ik_y}\\
-e^{ik_y} & -2\cos (k_x+2\pi\alpha) & -e^{-ik_y} & \ldots & 0 & i\Delta e^{ik_y} & -2i\Delta\cos
(k_x+2\pi\alpha) & i\Delta e^{-ik_y} & 0 & \ldots   \\
0 & -e^{ik_y} &  -2\cos (k_x+4\pi\alpha) & \ldots& & 0 & i\Delta e^{ik_y} & \ldots & \ldots &\dots \\
\vdots & & \ddots& & &\vdots & &\ddots& \\
2i\Delta\cos k_x & -i\Delta e^{-ik_y} & 0 & \ldots & -i\Delta e^{ik_y} &2\cos k_x & e^{-ik_y} & 0 &
\ldots & e^{ik_y}\\
-i\Delta e^{ik_y} & 2i\Delta\cos (k_x+2\pi\alpha) & -i\Delta e^{-ik_y} & 0 & \ldots &e^{ik_y} &
2\cos (k_x+2\pi\alpha) & e^{-ik_y} & \ldots & 0\\
0 & -i\Delta e^{ik_y} & \ldots & \ldots &\dots &0 & e^{ik_y} &  & \ldots\\
\vdots& & & & &\vdots
          \end{vmatrix}
\end{equation}
\end{widetext}
The gap satisfies the equation
\begin{eqnarray}
\Delta &= &\frac{V}{N}\sum_{\bf k}\left[\langle -ie^{ik_x} c^\dagger_{k_x+\pi,k_y+\pi} c_{k_x,k_y}
\rangle +\text{h. c.}\right]\nonumber \\
&=&\frac{V}{N}\sum_{\bf k} \Delta_{\bf k} .
\end{eqnarray}
For each value of $(k_x, k_y)$, the Hamiltonian was diagonalized, giving rise to a new set of fermion
operators $\{ d_{\bf k}^i\}$, $i= 1, \ldots, 2 q$. Then, $\Delta_{k_x,k_y}$,
$\Delta_{k_x+2\pi\alpha,k_y}$, \ldots , $\Delta_{k_x+2\pi\alpha(q-1),k_y}$,
$\Delta_{k_x+\pi,k_y+\pi}$, $\Delta_{k_x+\pi+2\pi\alpha,k_y+\pi}$, \ldots ,
$\Delta_{k_x+\pi+2\pi\alpha(q-1),k_y+\pi}$ were computed and so was the energy spectrum. Several
values of $k_x$ and $k_y$ were chosen in the sub-Brillouin zone (sBZ) and finally the gap was obtained
as 
\begin{equation}
\Delta = \frac{1}{2 q N_x N_y}\sum_{{\bf k}\in \text{sBZ}} \Delta_{\bf k},
\end{equation}
where $N=(qN_x)(2N_y)$. The whole process was repeated
until $\Delta$ converged.

 Typical magnetic fields in laboratories are of order
$20\,T$, equivalent to $(1/1225)\phi_0$ for the cuprates with lattice spacing of $4$ \AA.
In our computation, we were able to diagonalize matrices of size $4000\times 4000$, or
$q=2000$ corresponding to the smallest field of $12.5\,T$. Since the gap is so insensitive to the
applied field, going to smaller fields appeared unnecessary.
\begin{figure}[htb]
\includegraphics[scale=0.35]{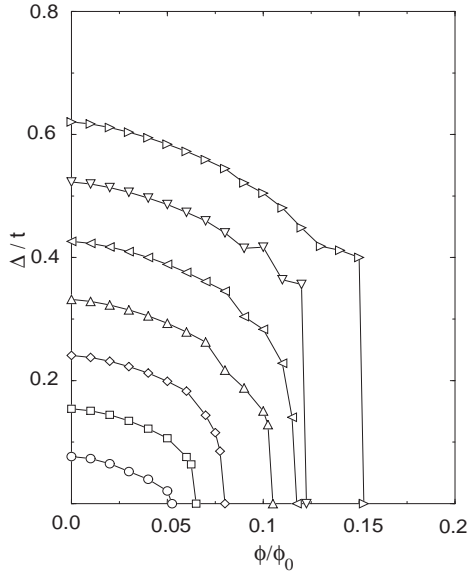}
\caption{The DDW gap magnitude $\Delta$ as a function of the applied flux
$\phi/\phi_0$ for various values of the coupling strength $V/t$ ($\bigcirc : 0.5$; $\Box : 0.75$;
$\Diamond: 1.0$; $\bigtriangleup : 1.25$; $\triangleleft : 1.5$; $\bigtriangledown : 1.75$;
$\triangleright : 2.0$)  for
$\mu =-0.1 t$. 
\label{fig:gap}}
\end{figure}

\begin{figure}[htb]
\includegraphics[scale=0.40]{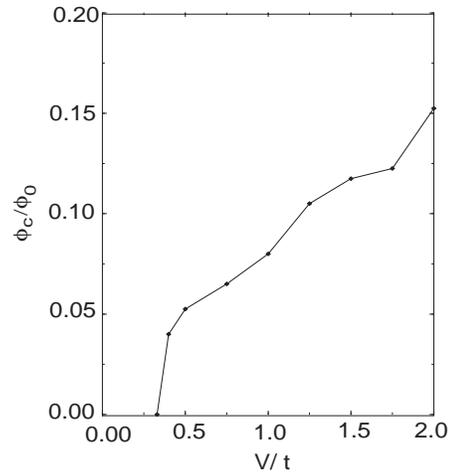}
\caption{The critical field as a function of $V$ for $\mu=-0.1 t$.
\label{fig:phic}}
\end{figure}

In Fig.~\ref{fig:gap}, we show the dependence of the DDW gap versus magnetic field at absolute zero 
for $\mu=-0.1 t$ as a function of $V$.  We observe a striking
insensitivity of the gap with respect to the magnetic field. The gap only collapses, through a
first-order transition at a flux strength $\phi_c/\phi_0$, corresponding to a large laboratory
magnetic field ($\phi_0\approx 25,000$T for cuprates). The gaps at zero field was calculated
separately, and the results for finite field were seen to converge to them as $\phi\to 0$.
It is important to note that because the chemical potential is finite, the DDW transition does not
take place at $V=0^+$, but at a finite threshold value $V_c$. This is clearly seen
in Fig.~\ref{fig:phic}. 
In the same figure, we see that $\phi_c/\phi_0$ drops quite rapidly to zero as $V_c$ is approached.
Thus, although $\phi_c$ corresponds to quite large magnetic fields when the DDW gap is large in the
underdoped regime, it could be quite small when the gap is small in the overdoped regime. For a gap of
order 20 meV, the critical magnetic field is of order 1250 T.

\begin{figure}[htb]
\includegraphics[scale=0.35]{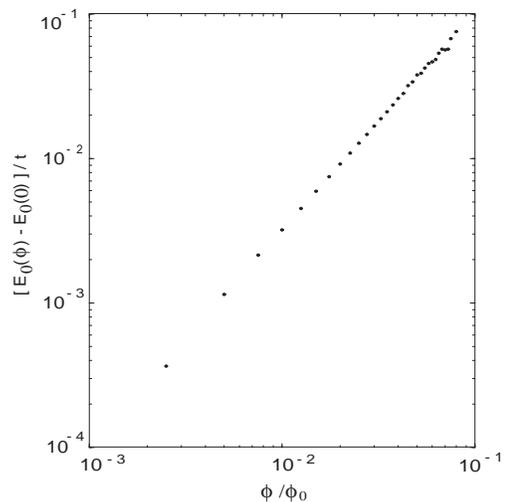}
\caption{The groundstate energy as a function of the applied magnetic field at $\mu = 0$.
\label{fig:E}}
\end{figure}

In order to test the correctness of our computations, we also calculated the 
groundstate energy density at half-filling ($\mu =0$). We observe an increase
in the groundstate energy as the magnetic field is applied; this increase indicates a
diamagnetism in the DDW system. This result has, in fact, been analytically derived 
\cite{Nersesyan} using the nodal fermi gas formalism of the DDW state for $\mu = 0$. In particular,
it was shown  that the DDW groundstate energy, expressed in terms of
lattice parameters, behaves as 
\begin{equation}
E_{GS}(\phi)=E_{GS}(0)+2\frac{\zeta(3/2)}{\sqrt\pi}\,A\,  t\,\sqrt{\frac{W}{t}}\left(
\frac{\phi}{\phi_0} 
\right) ^{3/2} ,
\label{eq:Nersesyan}
\end{equation}
where $\zeta(x)$ is Riemann's zeta function, $A$ and $W$ being the system 
area and a cutoff energy. The above formula is expected to hold only for small fields
such that $\hbar \omega_c=\hbar eB/m^*c<<W$. This appears to be a surprisingly robust result, as
shown in Fig.~\ref{fig:E}, especially because we made no approximations involving nodal Fermions.
The fit at the smallest field gave a sensible $W=1.6 t$.  The quantity $W$ was interpreted in
Ref.~\onlinecite{Nersesyan} as the cutoff energy 
below which the Dirac fermion description holds for the DDW state. 

As a consequence of the nonanalytic field dependence of the energy at $\mu=0$, 
the diamagnetic susceptibility diverges at zero field: $\chi \propto B^{-1/2}$. Obviously,
this divergence poses a serious question about the stability of the DDW state at $\mu =0$. It is
clear, however, that this divergence will be cut off for a finite chemical potential, $\mu\ne 0$.
Similarly, the divergence will also be removed at finite temperatures even for $\mu =0$. Moreover,
three dimensional coupling between the layers will also modify the result for the
susceptibility.\cite{Nersesyan} It is important to stress that the divergence in the susceptibility is
intimately related  to the appearance of zero-energy modes, which is a direct consequence of the nodal
points and the  doubling of the  Dirac fermions. 

It has recently
been proposed in
Ref.~\onlinecite{Balatsky} that a finite magnetic field may induce an additional  $d_{xy}$ gap. Such a
gap can efficiently remove the nodes and hence the divergence in the diamagnetic susceptibility. However,
the small 
$d_{xy}$ gap will not be manifest at a finite temperature and the dominant cutoff will be provided
by the temperature or the chemical potential. Therefore our computation for the
DDW gap retains its relevance to experimental situations.

As an interesting contrast to  DSC, an essential feature of the Dirac 
fermion picture of the DDW is the formation of Landau levels in a magnetic field:\cite{Nersesyan}
$E_n=\sqrt{n (\hbar \omega_c) \Delta_0}$, where $\Delta_0$ is the amplitude of the gap, and
$\omega_c$ is the cyclotron frequency, $eB/m^*c$, $m^*$ being the effective mass corresponding to the
nodal region. As a result, the DDW system away from half-filling is expected to exhibit  de Haas-van
Alfven (dHvA) oscillations in the magnetization. 

For a free electron system at chemical potential
$\mu$, where the energy spacing is
$\hbar \omega_c$, the oscillation frequency satisfies 
\begin{equation}\frac{\Delta B}{B}=-B\Delta\left(\frac{1}{B}\right)=\frac{\hbar\omega_c}{|\mu|}.
\end{equation}
Taking $B=1\,T$, $\mu\approx 0.25\,eV$, one would get $\Delta B \approx 5\,G$.
For nodal fermi gas at chemical potential $\mu$, the oscillation frequency is
\begin{equation} \frac{\Delta
B}{B}=-B\Delta\left(\frac{1}{B}\right)=(\frac{\hbar\omega_c}{|\mu|})(\frac{\Delta_0}{|\mu|}).
\end{equation} 
 To make an estimate, we can choose
$B=1\,T$. It is more problematic to estimate $\mu$ in the underdoped regime in which the
DDW state is likely to be found, and in which the typical non-interacting band picture may not hold.
There is some evidence, however, that 
$\mu\approx -t x^2$, where $x$ is the doping.\cite{Sumanta}  Choosing $x\approx 0.15$, 
$\Delta_0$ of the order of the pseudogap $\sim 20$ meV, and $t\approx 0.25$ eV,  results in
$\Delta B\approx 0.05$ T, which is clearly within the detectable range of dHvA oscillation
experiments.

The above estimate for $\Delta B$ is not sufficient for the detectability of dHvA effect. We must
also estimate the magnitude of the magnetization oscillations. For this purpose, it is convenient
to consider energy spectrum of the nodal  fermions of the DDW state.
\begin{figure}[htb]
\includegraphics[scale=0.35]{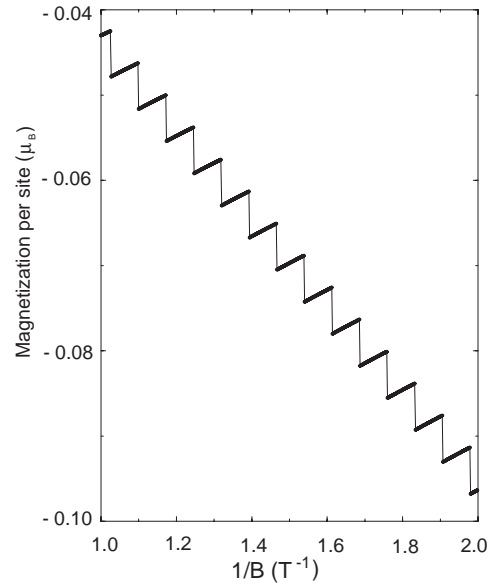}
\caption{The de Haas-van Alphen oscillations of the magnetization in the DDW state.
\label{fig:dHvA}}
\end{figure}
The oscillation in energy, and thus of the magnetization, occurs
whenever a Landau level passes through the Fermi level. For  hole-doped cuprates, with a finite
chemical potential $\mu$,   the formula (\ref{eq:Nersesyan}) for the groundstate energy must be
augmented by the following additional term:
\begin{equation}
\frac{eB}{hc}\sqrt{\hbar\omega_c\Delta_0}\ \sum_{n=0}^{n_\mu}\sqrt{n}
\end{equation}
where $[n_\mu]$ is the largest integer smaller than $\mu^2/(\hbar \omega_c \Delta_0)$. 
In other words, $n_\mu$ is the number of filled Landau levels. With the above choice of
parameters, the  magnetization per unit cell, in units of Bohr magneton, $\mu_B$, is
shown in Fig.~\ref{fig:dHvA}. In plotting this figure, we have removed the spurious
$\delta$-function singularities arising from taking derivatives of the step functions. At finite
temperature, the sharpness of the oscillations will be smoothed out,  but the oscillation amplitude
will still roughly be of order $\approx 0.01\,\mu_B$, which is  quite weak. As is evident from
Fig.~\ref{fig:dHvA}, as the magnetic field swept between
$0.5$ T and $1$ T, the system undergoes about $14$ oscillations.

There are already some experiments that appear to shed  light on the effect of  magnetic field
on the  pseudogap. Nuclear magnetic resonance experiments in underdoped
$\mathrm{YBa_2Cu_4O_8}$ clearly support the view that the pseudogap is hardly affected by an
applied perpendicular magnetic field,\cite{Zheng} and so does experimets in nearly optimally
doped\cite{Pennington}
$\mathrm{YBa_2Cu_3O_{7-\delta}}$. In contrast, nuclear magnetic resonance experiments in
overdoped $\mathrm{TlSr_2CaCu_2O_{6.8}}$ provides contrasting evidence that in this regime the pseudogap
is sensitive to the magnetic field.\cite{Zheng2} At first sight, the experiments in
$\mathrm{YBa_2Cu_3O_{7-\delta}}$ appear to be controversial because there are also reports to the
contrary,\cite{NMR2,NMR3} but as was pointed out by the authors of Ref.~\onlinecite{Zheng2}, the later
reports by the authors of Ref.~\onlinecite{NMR2,NMR3} seem to be in agreement with
Ref.~\onlinecite{Pennington}. Therefore, our theoretical results would be consistent with  experiments,
if the pseudogap is indeed caused by DDW. We hope that our work will catalyse further experiments on this
important question. On the theoretical front, it will be necessary to go beyond the Hartree-Fock
approximation, especially at finite temperatures. To go beyond Hartree-Fock approximations, we must
recognize the fact that the DDW gap, which characterizes the orbital currents need not be in the perfect
staggered pattern, but can be  flipped as long as the current conservations at the vertices are
satisfied.\cite{Chakravarty} We hope to report on this issue in the future.

We thank A. V. Balatsky, Pengcheng Dai, H. A. Mook, and C. Nayak for interesting discussions. This work
was supported by a grant from the National Science Foundation: NSF-DMR-9971138.

\end{document}